# Exploration of the Doping Effect in the Thiolate-protected Gold Nanoclusters: DFT Simulations of H2S-nanoalloy Complexes

Hui Jia, Changlong Liu and Yonghui Li[*]

Department of Physics and Tianjin Key Laboratory of Low Dimensional Materials Physics and Preparing Technology, School of Science, Tianjin University, Tianjin 300350, China

Email: yonghui.li@tju.edu.cn

## Abstract

The atomically precise method has become an important technique to adjust the core of thiolate-protected gold nanoclusters to improve physical and chemical properties. But the doping effect on the structural stability has not been systematically summarized. In this work, the $H_2S$-nanoalloy molecules with different doping metal atoms has been investigated to elucidate the impact of the dopant on the structures. With DFT simulation results, the zinc group atoms as dopants may be influenced by surrounded gold atoms and the binding of the thiolate units are enhanced. The simulated zinc group data when combined to the gold group and plantinum group data can be summarized in the perspective of balance between the ligand-core binding and core cohesive energies. Most of dopants drive the modeled nanoclusters away from the balance especially when the metal atom replaced the gold atom in gold-sulfur bindings. But when cores of the nanoclusters are dominated by gold atoms, the dopants may achieve "saturation" such that the balance in the doped clusters may be corrected. This work provide a simple profile to understand the internal shift of the structure introduced by the atomically precise method.

## I. Introduction

Coinage metal nanoalloys, especially gold-contained nanoalloys, have drawn considerable attraction in the past decades for their structural and electrical properties.[1-6] Among their special properties, the molecular chemisorptions on small nanoalloys have become more and more important.[7-10] This is partly because of the continuous experimental and theoretical work on metallic nanoclusters. The discovery of the thiolate-protected gold nanoclusters promotes the importance of molecular chemisorptions to a more important level. Adhesions between thiolate ligands and gold cores[11-13] seem to follow a similar rule as the molecular chemisorptions of coinage metal.

As the emergence of atomically precise synthesis brings the adjustability to the cores of gold nanoclusters, "chemically doped" metallic atoms in the cores may bring significant improvements in optical or catalytical properties.[14-19] Such versatility also introduces extra complications of the interaction between the ligands and the cores. To understand such interactions, molecular chemisorption of small coinage nanoalloys (nanoclusters) are interesting and enlightening models to study. With the models with $H_2S$ molecules adsorbed, the reported bond modulations and stiffness of bindings can be systematically investigated with more intuitive interpretations of "doping effects".

Recent developments with theoretical research of molecular chemisorptions of small coinage nanoclusters have partially depicted "doping effects" which brings by the guest atoms into nanoclusters. Pakiari studied hydrogen sulfide and the atoms of the gold group (Cu, Ag and Au) with DFT to investigate the properties of the sulfur-metal bond.[20] The nature of the bonding between gold nanoclusters doped by platinum group atoms (Ni, Pd, and Pt) and hydrogen sulfide molecules was simulated by Ge.[21] Different groups of doping atoms are systematically explored with more details provided intuitively but the doping effect has not summarized and discussed. As summaried in the previous work [22], the internal balance inside each nanocluster has been pointed out. It is also important to review the doping effect introduced by a dopant in a lightly doped gold nanocluster, especially in the perspective of balance between ligands and the core.

In this work, after the simulation methods are summarized in section II, the simulation results of modeled systems with doping of zinc group atoms are analyzed and discussed in section III. Then in section IV, global rules across different chemical groups are summarized to give some insights of doping effect followed by the summation in the last section.

## II. Simulation Method

To investigate the impact of the doped atom on the $H_2S$-nanocluster complexes, we limit the nanoclusters in our simulations to up to 3 metallic atoms. Despite the details of the doping in the larger clusters such as $Au_{25}(SCH_2CH_2Ph)_{18}$[23], $H_2S$-nanocluster complexes resemble the thiolate-protected gold nanoclusters with doped atoms intuitively.

Density functional theory (DFT), which is a "formally exact" theory to simulate the electronic density in many-body systems, has been applied in many disciplines including physics, chemistry, and biology. As the exchange-correlation potential (Vxc) in the theory is well approximated, the simulated electronic density and density-related quantities are guaranteed to be the exact ones. To get a common trend that does not depend on the approximations, we selected 6 different Vxc functionals for simulations including B3LYP [24], CAM-B3LYP [25], PBE [26], PBE0 [27], PW91 [28] and PW91-P86. The models in this work are simulated with widely used software in quantum chemistry, Gaussian 09 [29], which provides a variety of basis sets and Vxc functionals. B3LYP (Becke, 3-parameter, Lee-Yang-Parr hybrid functionals) functional is used in simulations which contain zinc group dopants. CAM-B3LYP functional adds long-range corrections to the B3LYP functional can help in showing the importance of long-range interactions. PBE (Perdew–Burke–Ernzerhof) functional is the most successful generalized gradient approximation (GGA) which has been used in many solid-state applications. Its variant, PBE0 functional allows more than just density-based features by adding a 25% exact exchange. The first reasonable GGA that can be reliably used over a very wide range of materials named PW91 (Perdew-Wang 1991, the precursor of PBE) functional is also in the scope of our consideration. The modified version of PW91 by replacing its correlation part by its older variance PW86 is also used in this work. To keep the precision of the simulations, the basis set for S, H and metals are 6-311++G**, 6-31G and aug-cc-pVDZ-PP [30] respectively. The diffusion component in the basis set is critical in the capture of the long-range interaction. For the simulations with dopants across zinc, gold and plantinum groups, B3LYP is the only Vxc to reduce the computational cost.

The doping effects and the metal-sulfur interactions can be evaluated based on the simulated density. One can calculate the binding energies for any two parts as

$$E_b = E_{total} - E_{part1} - E_{part2} \quad (1).$$

Such definition applies to arbitrary split into two parts. The absolute value of binding energy (or the value of dissociation energy, respectively) represents the firmness of binding.

Besides, the Fukui function method is also simulated to qualitatively analyze the relative activity within a chemical molecule by adding or subtracting a certain number of electrons. The magnitude of the Fukui function at a particular site represents its reactivity. The Fukui function is defined as

$$f(\vec{r}) = \left(\frac{\delta \mu}{\delta v(\vec{r})}\right)_N \quad (2).$$

where $\mu$ is the chemical potential, $v(\vec{r})$ is the external potential, and the derivative is obtained with a constant number of electrons N. [31] In practice, the defined Fukui functions are not exactly valued but approximated by the electrophilic Fukui equation or the nucleophilic Fukui equation in differential forms [32]:

$$f^- = n_N(\vec{r}) - n_{N-1}(\vec{r}), (3)$$
$$f^+ = n_{N+1}(\vec{r}) - n_N(\vec{r}). (4)$$

$n_N(\vec{r})$ is the electron density of N electrons out of the simulation.

The Natural Population Analysis (NPA) can be performed based on the DFT simulation results. The critical step in NPA is the Natural Bond Orbitals (NBO) Transformation of Natural Orbitals which are obtained from the diagonalization from the 1 particle reduced density matrix from DFT. NPA is used to calculate the number of orbital populations of atomic charges and molecular wave functions in general atomic orbital basis groups. This method can be exhibited with higher numerical stability and better describe the electron distribution in compounds.

In addition to Fukui functions and NPA, the ligand-core binding energy (BE) and core cohesion energy (CE) defined in the previous work[22] provide more insites to the structure information of the nanoclusters:

$$\text{Ligand} - \text{core binding enenrgy} = \frac{E_{Full\ complexe} - E_{Ligand} - E_{Core}}{N_{Ligand\ metal}}, \quad (5)$$

$$\text{Core cohesion energy} = \frac{E_{Full\ complexe} - E_{Ligand} - N_C E_{Metal\ atom}}{N_C + N_{Ligand\ metal}}. \quad (6)$$

Where $N_C$ represents the number of metal atoms in the core part and $N_{Ligand\ metal}$ represents the number of metal atoms in the ligand part. To classify a metal atom to core or ligand, a simple rule which is slightly different from the literature is used: an atom is a part of core/ligand when its NPA charge is negative/positive. The ligand part includes the hydrogen sulfide molecule. Such an algorithm allows us to explore the metal atoms from different families over the periodic table.

## III. Structural and Interaction Characters of Zinc Group Dopants

A series of zinc group related small molecules are simulated including alloy binary clusters $M_1$-$M_2$ ($M_{1,2}$=Au, Cd, Hg, and Zn), $H_2S$-M, $H_2S$-MAu, $H_2S$-AuM, $H_2S$-$Au_2$M and $H_2S$-$MAu_2$ molecules (M=Au, Cd, Hg and Zn). The interactions between zinc group atoms in binary clusters are very weak. The binding energies in the binary clusters without gold atoms are weaker by one order of magnitude than in the clusters with gold atoms (Table S1). The simulated vibrational frequency which corresponds to the stiffness in harmonics behaves similarly in the trend of the binding energies. The binary clusters agrees the electronic structures of the zinc group metals which tend to form the closure shells rather than to form bonds with other atoms. Therefore, it is difficult for the atoms in zinc group metals to form clusters.

On the other hand, the stable electronic structure of zinc group atoms may be break by large affinity of gold atoms. Allowing one gold atom in binary clusters significantly enhance the stability of the binary alloy in the binding. Better enhancement which inclds a shorter bond length, lower binding energy, and higher vibrational frequency are seen in Au-M clusters with smaller zinc group metal. Such impact of gold is demonstrated in the upper panel *Figure 1*a: zinc group dimers show clear sparation from other dimers. But when one atom in the dimer is replaced by a gold atom, the separation disappears while all the zinc group Au-M clusters distribute in the region of low binding energy. In addition, the simulated oscillation frequencies of all the pure dimers are negatively proportional to the bond lengths as:

$$\omega(r) = -358.3r + 1083.3 \quad (7)$$

with score ($R^2$) 95% which is shown in *Figure 1*b. The Au and Cd atoms lies in the long bond length end which indicates their poor bonding potential.

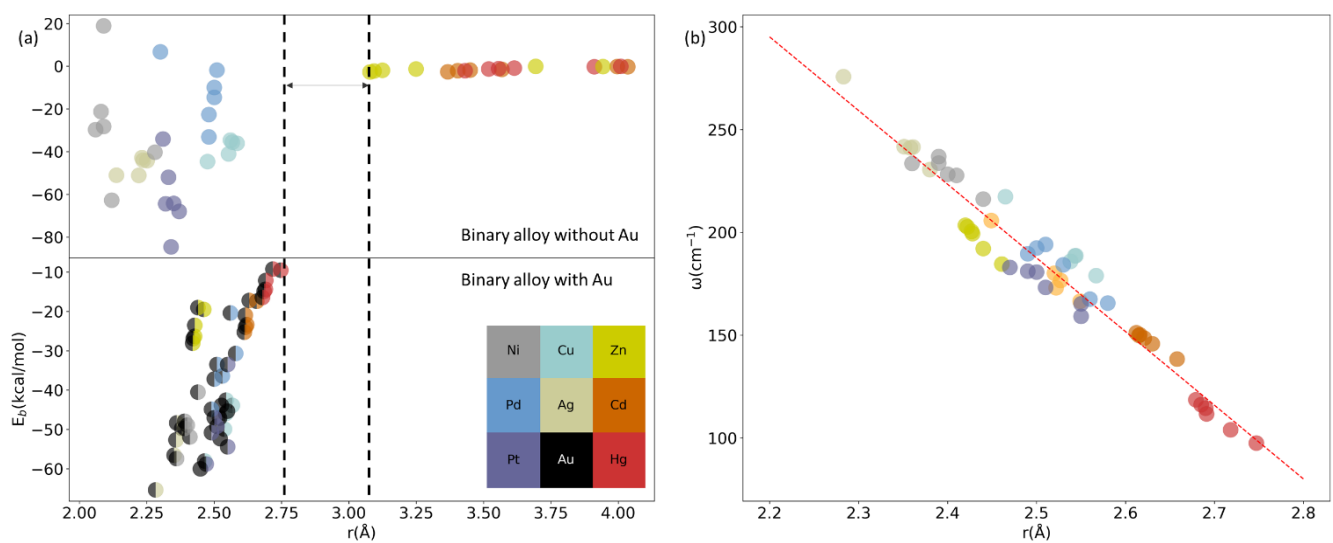

*Figure 1 (a) "separation" between the zinc group dimers and other dimers which can be shifted by substituting one of the metal atom in each dimer by a gold atom. Gold substitution impacts the bond length more significantly than the binding*

*energy. (b) Linear correspondency ($R^2=0.95$) between harmonic frequencies and bond lengths of binary alloy dimers.*

With the general information from the binary alloies, $H_2S$-cluster molecules are simulated as the smallest thiolate-protected doped gold nanoclusters to investigate the interactions between the ligand (represented by $H_2S$) and the doped gold core (represented by doped nano alloy). As summarized in *Table 1***Error! Reference source not found.**, the molecules with zinc group dopants show extended bond lengths, reduced the vibration frequencies, and decreased the binding energies. Among the doped molecules, structures in which Zn/Cd atom plays as the attaching spots are more energetically favored than other possibilities. With the Zn/Cd attaching spots, other two gold atoms can significantly improve the metal-sulfur interactions**Error! Reference source not found.** by reducing their bond lengths, increasing vibrational frequency, and reducing binding energies (about 10eV). The Cd doped gold nanocluster is also reported in the literature [33]. But when a Hg atom serves as the attaching spot, other two gold atoms fail in improving the interaction between Hg and S. The Hg-S binding is still weak such that the gold atoms approach to the hydrogen atoms in the relaxation procedure.

*Table 1 Selected features include metal-sulfur distances (r in Å), metal-sulfur vibrational frequencies (ω in $cm^{-1}$), and binding energies ($E_b$ in kcal/mol) of $H_2S$-dimer, trimer complexes.*

| $V_{xc}$ | molecules | r(S-M) | ω (S-M) | $E_b$ | molecules | r(S-M) | ω (S-M) | $E_b$ |
|---|---|---|---|---|---|---|---|---|
| PBE | 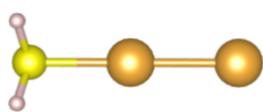 | 2.322 | 278.2 | -26.37 | 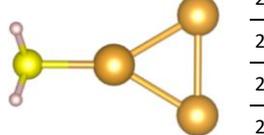 | 2.304 | 285.6 | -26.82 |
| PBE0 | | 2.330 | 271.9 | -23.63 | | 2.310 | 281.9 | -26.6 |
| CAM-B3LYP | | 2.360 | 259.8 | -21.64 | | 2.341 | 268.7 | -23.44 |
| B3LYP | | 2.381 | 244.3 | -19.33 | | 2.360 | 253.9 | -21.54 |
| PW91P86 | | 2.321 | 281.3 | -28.76 | | 2.304 | 288.0 | -31.05 |
| PW91PW91 | | 2.322 | 278.7 | -26.87 | | 2.305 | 286.1 | -29.19 |
| PBE | 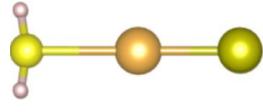 | 2.609 | 133.0 | -7.74 | 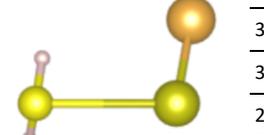 | 2.970 | 46.5 | -3.23 |
| PBE0 | | 2.595 | 134.4 | -6.54 | | | | |
| CAM-B3LYP | | 2.636 | 129.1 | -5.23 | | 3.318 | 46.5 | -1.02 |
| B3LYP | | 2.732 | 110.7 | -3.75 | | 3.294 | 48.9 | -0.75 |
| PW91P86 | | 2.599 | 138.0 | -9.59 | | 2.936 | 91.7 | -4.79 |
| PW91PW91 | | 2.607 | 134.5 | -8.17 | | | | |
| PBE | 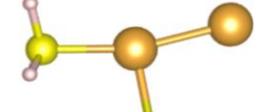 | 2.336 | 270.0 | -9.63 | 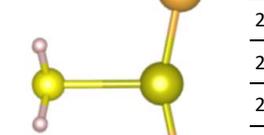 | 2.675 | 66.0 | -16.80 |
| PBE0 | | 2.347 | 262.9 | -8.30 | | 2.682 | 49.4 | -17.26 |
| CAM-B3LYP | | 2.386 | 246.2 | -6.21 | | 2.732 | 68.8 | -14.06 |
| B3LYP | | 2.421 | 221.7 | -4.03 | | 2.802 | 65.7 | -13.10 |
| PW91P86 | | 2.334 | 273.6 | -11.65 | | 2.644 | 65.8 | -17.82 |
| PW91PW91 | | 2.337 | 269.9 | -10.04 | | 2.664 | 66.1 | -17.16 |
| PBE | 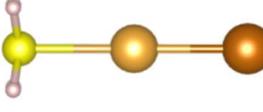 | 2.586 | 138.2 | -7.86 | 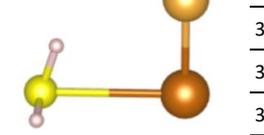 | 3.283 | 41.7 | -3.10 |
| PBE0 | | 2.576 | 144.2 | -6.58 | | 3.307 | 30.2 | -2.15 |
| CAM-B3LYP | | 2.619 | 144.1 | -5.27 | | 3.508 | 23.4 | -1.05 |
| B3LYP | | 2.714 | 132.3 | -3.74 | | 3.535 | 23.4 | -0.71 |
| PW91P86 | | 2.574 | 142.0 | -9.83 | | 3.210 | 43.8 | -4.71 |
| PW91PW91 | | 2.583 | 140.0 | -8.33 | | 3.255 | 44.4 | -3.44 |
| PBE | 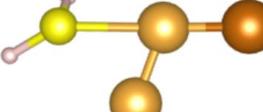 | 2.330 | 272.8 | -11.29 | 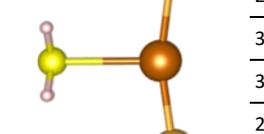 | 2.978 | 31.8 | -10.92 |
| PBE0 | | 2.337 | 269.0 | -9.95 | | 2.981 | 34.2 | -12.04 |
| CAM-B3LYP | | 2.375 | 251.6 | -7.95 | | 3.033 | 29.8 | -10.24 |
| B3LYP | | 2.404 | 232.9 | -5.54 | | 3.108 | 27.0 | -8.55 |
| PW91P86 | | 2.328 | 277.5 | -13.46 | | 2.932 | 31.1 | -11.99 |
| PW91PW91 | | 2.331 | 273.4 | -11.67 | | 2.962 | 32.2 | -11.26 |
| PBE | | 2.525 | 148.9 | -9.89 | | | | |
| PBE0 | | 2.524 | 150.4 | -7.84 | | 3.724 | 46.0 | -1.17 |

| Functional | | | | | | | |
|---|---|---|---|---|---|---|---|
| CAM-B3LYP | | 2.574 | 141.5 | -6.06 | | | |
| B3LYP | | 2.653 | 125.6 | -4.92 | 3.828 | 31.7 | -0.06 |
| PW91P86 | | 2.517 | 155.7 | -12.00 | 3.301 | 66.1 | -2.70 |
| PW91PW91 | | 2.524 | 151.2 | -10.37 | | | |
| PBE | | | | | | | |
| PBE0 | | | | | | | |
| CAM-B3LYP | | 2.370 | 256.6 | -10.97 | 3.463 | 37.7 | 3.83 |
| B3LYP | | 2.392 | 239.9 | -8.98 | 3.566 | 35.2 | 4.59 |
| PW91P86 | | 2.327 | 277.1 | -16.59 | 3.841 | 26.0 | -3.99 |
| PW91PW91 | | | | | 3.078 | 40.2 | -4.1 |

To capture the details of the metal-sulfur interaction, 6 different exchange-correlation functionals are used and our discovery does not depend on a particular functional. The long-range correction of B3LYP indicates that the B3LYP functional may underestimate the bond length and over estimate the binding energies. The exchange component in the functionals seems to be more important. In mecuray doped cases, PW91 seems to be a good functional to use even its correlation part may be changed to its older form.

To understand the metal-sulfur binding in the model systems, the donor-acceptor model is applied based on the results of NBO analysis (*Figure 2*a, b). In the H$_2$S-M molecules the charge transfer locate between hydrogen and sulfur atoms which agrees the explanation of weak interaction explained. As gold atoms are added to the molecule and the molecules are extended into H$_2$S-MAu$_n$ (n=1,2), zinc group atoms are forced to transfer electrons to gold atoms (*Figure 2*c). As a result, the zinc group atoms begin to show "induced" affinity toward sulfur atoms which allow the "ligand-core" (metal-sulfur) binding raise to a significant level.

The NBO analysis (Table S3) shows more details: the 4s electron of the Zn atom is transferred to its 4p orbital, the 5s electron of the Cd atom is also transferred to its 5p orbital, and the 6s electron of the Hg atom is shifted to the 6p orbital. Such sp hybrid characters are strongly directional, which may be the source of the enhancement of the interaction.

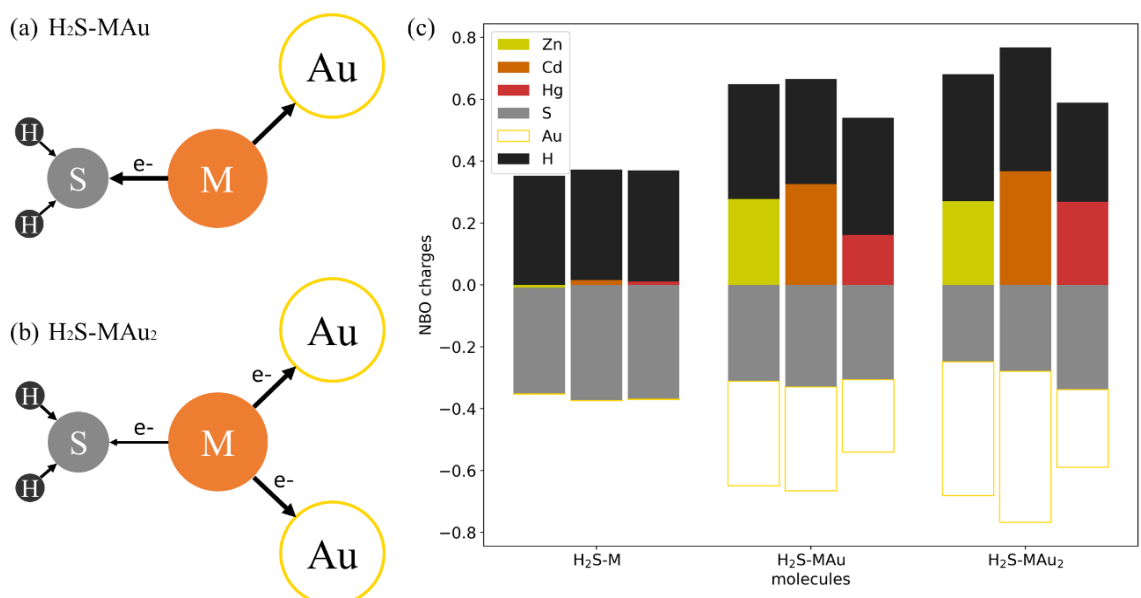

*Figure 2 Charge transfer of doped clusters in (a) H$_2$S-MAu and (b) H$_2$S-MAu$_2$ molecules. The black arrows represent the charge transfer direction indicated by the natural electron configuration and natural charge. (c) NBO analysis results are summarized in a bar chart. The electrons are transferred from hydrogen and metal atoms to sulfur and gold atoms. As the number of gold cluster increases, extra electrons are removed from metal, sulfur and hydrogen atoms.*

The electrophilic Fukui function ($f^-$ defined as equation 3) of the simulated molecules are shown in *Figure 3*. Smaller isosurfaces of the Fukui function represent the weaker chemical activity. The molecules in *Figure 3*d-f respectively correspond to the molecules in *Figure 3*a-c each with an extra atom. The results indicate that the electrophilic active site is shifted from the doped metallic atoms (Zn, Cd and Hg) to the extra gold atoms and the electrophilic activity of the sulfur atom in the hydrogen sulfide molecule is reduced. The doped metallic atom (Zn, Cd, or Hg) and the metal-sulfur bond are protected by the doped gold atoms. The adding gold atom also affects the zinc group atom in the molecule when the zinc group metal serves as the adhesion spot as shown in **Error! Reference source not found.**g-l. The electrophilic active region is almost totally transferred from the zinc group atom to the added gold atom. The Fukui function indicates that the doped Zn, Cd or Hg atoms may be chemically saturated by two gold atoms. Such saturations are localized between the $H_2S$ units and metallic atoms but further aggregation along the active gold atoms is also possible. Whether the metal atom (Zn, Cd, or Hg) is the central metal atom or not, the extra doped gold atom can play the role of protector. This makes the doped clusters more stable and beneficial for cluster developments. Such results also indicate that the reactivity of the zinc group atoms in the thiolated-protected nano cluster should be more than stationary when the molecular dynamics is required [33].

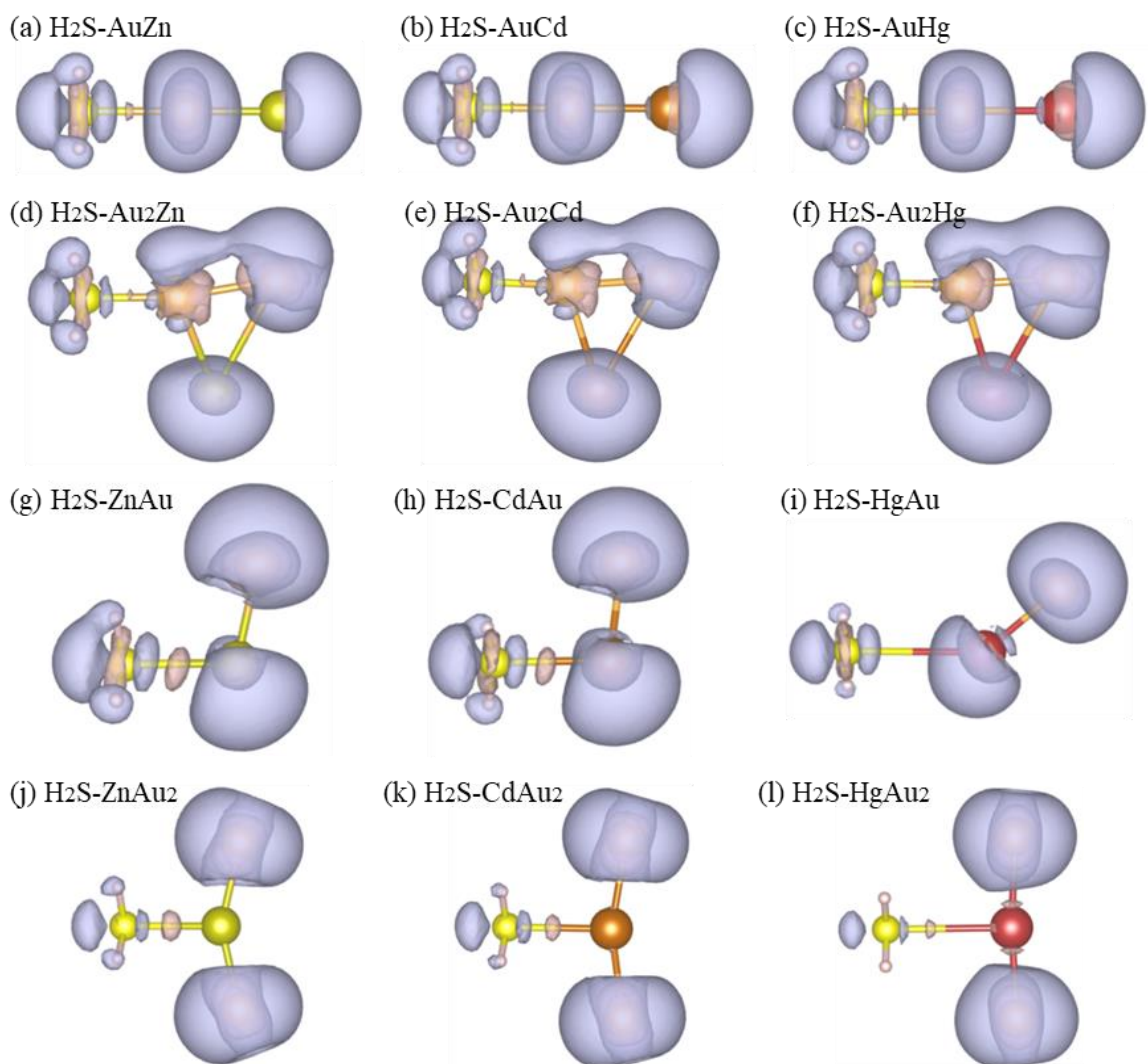

*Figure 3 Fukui function of (a)-(f) $H_2S-Au_nM(n=1-2)$ and (g)-(l) $H_2S-MAu_n(n=1-2)$. The purple and cyan colors correspond to the areas with positive and negative isovalue surfaces, respectively.*

## IV. Doping Impacts on the Balace between the Ligand-core Binding and Core Cohesion

Considering the balance between BEs and CEs, most undoped ligand protected gold nanoclusters falls into the equilibrium status. But as the single atom dopings are introduced to the molecules, the equilibria may be attenuated. To understand such attenuation effect, the simulation results including zinc group elements (this work), Copper group elements (reference 20) and platinum group elements (reference 21) dopant are summarized in *Figure 4*. The equilibria between BEs and CEs are represented by the black solid line. In the doped clusters, most dopants belongs to ligands for their accumulation of positive charges indicated by NBO analysis. BE-CE balance is a different scope beyond the $H_2S$ binding energy to understand the modeled cluster. The results show that most of the doped molecules lie below the equilibria line which indicates that the dopant atoms breaks the equilibria by reducing core cohesive and/or enhance the ligand-core binding.

Though the dopants attenuates the BE-CE balance, some of the unbalance BE-CEs can be partially corrected by adding more gold atoms to the core. As a dopant reaches its saturation of loosing electrons among gold atoms, adding more gold atoms improves CE only. If we allow 10 eV deviation from the line of equilibria which is represented by the dotted line in *Figure 4*, half of the large molecules (with >4 metal atoms) lies within the range lightly deviation. By comparing the molecules with core-doping and ligand-doing, it is more stability favored for the clusters with core-doping.

Doped clusters behaves similarly with a few exceptions including $H_2S$-$CdAu_n$ and $H_2S$-$CdAu_n$. Analysis of NPA results indicate that doping gold atoms to some level can enhance the stability in $H_2S$-$CdAu_2$, but too many gold atoms increase the CE only. Compared to other metals the electron affinity of Cd is too weak and must be improved by gold atoms.

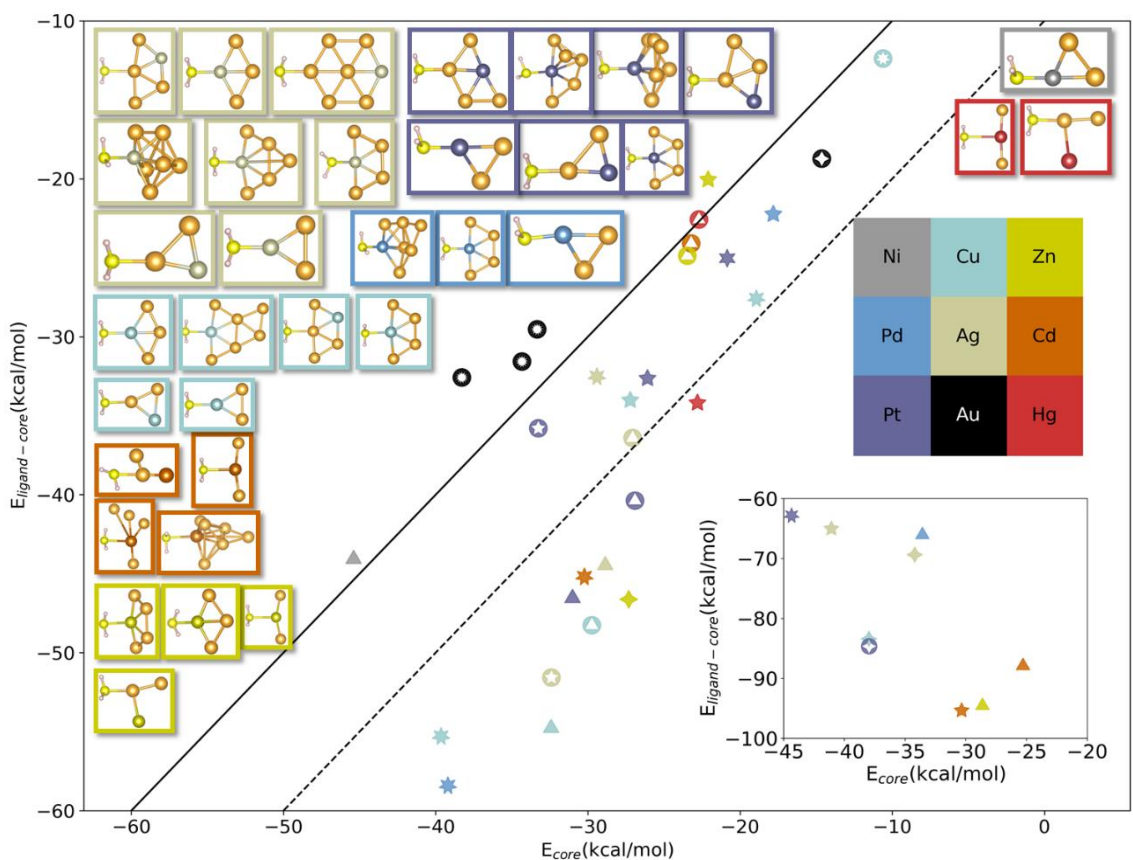

*Figure 4 Summary of the BE-CE balance for all modeled simple clusters H2S-MAun and H2S-AunM. The molecules with doped atoms follow the color table in the figure. H2S-MAun corresponds to the N+1-pointed star while H2S-AunM corresponds to cut N+1-pointed star. (M = Cu, Ag, Ni, Pt, Pd, Zn, Cd and Hg)*

## V. Summary

In this paper, the electron structure properties of zinc group atoms in dimer or H2S contained molecules are investigated and discussed. Simulation of dimer cluster cores reveals that the electronic structure of zinc group elements tends to form closed shells, so it is difficult for atoms of pure zinc group atoms to form clusters. At last, in the tri-group metals, the bond length and frequency of the alloy dimer are found to be inversely related. There is a separation in binding energies and bond lengths between zinc group atoms and other group atoms.

Adhesion of a hydrogen sulfide molecule is significantly affected by the doped gold atoms. The simulated results show that for the Zn atom and Cd atom, the addition of two gold atoms can significantly improve the metal-sulfur interactions. Then we used the donor-acceptor model to illustrate the binding of doped metal atoms and hydrogen sulfide molecules. The hydrogen sulfide molecule acts as a donor and the central metal atom (Zn, Cd, or Hg) acts as an acceptor. When the doped gold atom is given electrons of the central metal atoms, the central metal atoms have enough free orbitals to accommodate the electrons of the sulfur atom. To understand the active site of the complex, the Fukui function is calculated. It can be found that the electrophilic active site is displaced after additional gold atom doping. In addition, the electrophilic active region becomes smaller after doping.

When zinc group, gold group and plantinum group metals are summarized, a clear rule of dopants is shown. In the term of BE-CE balance, the dopants reduce the balance by significantly decreasing the BE and/or increase CE. As summarized in previous work [34], among ligand-protected gold nanoclusters, clusters with larger ligands are more stable in general.

# Appendix

### Structure and stability of binary alloy dimers

The simulation results of alloy binary clusters $M_1$-$M_2$ ($M_{1,2}$=Au, Cd, Hg, and Zn) and M-Au (M=Cd, Hg, and Zn) are shown in **Error! Reference source not found.**. The binding energies in the binary clusters without gold atoms are weaker by one order of magnitude than in the clusters with gold atoms. <span style="color:red">The simulated vibrational frequency behaves similarly in the trend of the binding energies.</span> Electronic structures of the zinc group metals tend to form the closure shells rather than to form bonds with other atoms. Therefore, it is difficult for the atoms from the pure zinc group metals to form clusters. However, gold atoms may break the stable shells of electrons of the zinc group atoms by their large affinity. As the atomic number increase, the Au-M binary cluster shows weaker binding including a longer bond length, higher binding energy, and lower vibrational frequency. Besides, the simulation results suggest that the spins prefer to keep the magnetic dipole as low as possible.

*Table S1 The bond lengths ($r_e$ in Å), vibrational frequencies ($\omega_e$ in cm-1), and binding energies ($E_b$ in kcal/mol) were calculated using different density functional theory methods including B3LYP, CAM-B3LYP, PBE, PBE0, PW91P86, PW91PW91.*

|  | Au-Au | | | Zn-Zn | | | Cd-Cd | | | Hg-Hg | | | Cd-Hg | | |
|---|---|---|---|---|---|---|---|---|---|---|---|---|---|---|---|
| Method | r | ω | $E_b$ | r | ω | $E_b$ | r | ω | $E_b$ | r | ω | $E_b$ | r | ω | $E_b$ |
| B3LYP | 2.549 | 166.5 | -45.35 | 3.693 | 18.0 | -0.03 | 3.996 | 17.8 | -0.17 | 4.009 | 12.0 | -0.05 | 4.011 | 13.7 | -0.12 |
| CAM-B3LYP | 2.527 | 176.7 | -43.63 | 3.943 | 17.2 | -0.11 | 4.035 | 20.2 | -0.25 | 3.911 | 18.1 | -0.27 | 3.952 | 14.7 | -0.28 |
| PW91PW91 | 2.521 | 173.3 | -53.46 | 3.093 | 55.0 | -2.2 | 3.404 | 41.4 | -2.09 | 3.519 | 29.4 | -1.32 | 3.436 | 33.0 | -1.73 |
| PW91P86 | 2.518 | 174.7 | -55.72 | 3.078 | 54.8 | -2.64 | 3.367 | 47.4 | -2.61 | 3.431 | 27.3 | -2.04 | 3.371 | 37.7 | -2.37 |
| PBE0 | 2.519 | 175.5 | -47.05 | 3.249 | 51.0 | -1.31 | 3.568 | 31.9 | -1.44 | 3.614 | 21.2 | -0.84 | 3.572 | 26.4 | -1.14 |
| PBE | 2.523 | 172.0 | -52.89 | 3.124 | 55.9 | -1.91 | 3.450 | 33.5 | -1.79 | 3.556 | 27.6 | -1.07 | 3.484 | 29.6 | -1.44 |
|  | Zn-Cd | | | Zn-Hg | | | Au-Zn | | | Au-Cd | | | Au-Hg | | |
| Method | r | ω | $E_b$ | r | ω | $E_b$ | r | ω | $E_b$ | r | ω | $E_b$ | r | ω | $E_b$ |
| B3LYP | 3.883 | 13.4 | -0.09 | 3.989 | 15.7 | -0.05 | 2.461 | 184.6 | -19.47 | 2.658 | 138.4 | -17.38 | 2.747 | 97.5 | -9.5 |
| CAM-B3LYP | 3.985 | 20.3 | -0.16 | 3.944 | 15.9 | -0.2 | 2.440 | 192.1 | -19.01 | 2.630 | 145.8 | -17.22 | 2.718 | 104.0 | -9.2 |
| PW91PW91 | 3.256 | 56.0 | -2.09 | 3.240 | 44.4 | -1.75 | 2.422 | 202.7 | -26.89 | 2.616 | 150.2 | -24.01 | 2.685 | 116.2 | -14.9 |
| PW91P86 | 3.234 | 57.0 | -2.58 | 3.193 | 46.0 | -2.39 | 2.420 | 203.5 | -28.08 | 2.612 | 151.2 | -25.25 | 2.679 | 118.6 | -16.5 |
| PBE0 | 3.401 | 35.8 | -1.35 | 3.391 | 32.2 | -1.07 | 2.428 | 199.4 | -23.49 | 2.615 | 149.9 | -20.96 | 2.691 | 111.7 | -12.1 |
| PBE | 3.284 | 53.3 | -1.8 | 3.277 | 42.5 | -1.46 | 2.427 | 200.5 | -26.31 | 2.621 | 148.6 | -23.39 | 2.690 | 114.5 | -14.4 |

### The structure and stability of $H_2S$-$M_n$ (n=1,2,3) complexes

In this section, we focus on the stability of the adhesion between ligands and cores of clusters. The modeled complexes are constituted of one of the metallic cores including metal atoms, binary clusters, and ternary clusters, and one attached hydrogen sulfide molecule on each core. The simulation results of $H_2S$-M models are shown in **Error! Reference source not found.**.

Different DFT functionals suggest that the metal-sulfur (M=Cd, Hg, and Zn) bindings are weak.

*Table S2 Selected features include the metal-sulfur distances ($r_e$ in Å), metal-sulfur vibrational frequencies ($\omega_e$ in cm-1), and binding energies ($E_b$ in kcal/mol) of the H$_2$S atomic complex.*

| Complexes | Method | r(S-M) | ω (S-M) | $E_b$ |
|---|---|---|---|---|
| 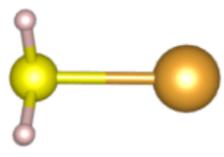 | PBE0 | 2.482 | 162.3 | -8.57 |
| | CAM-B3LYP | 2.553 | 141.5 | -6.08 |
| | PBE | 2.459 | 175.3 | -12.44 |
| | B3LYP | 2.598 | 132.9 | -5.95 |
| | PW91P86 | 2.453 | 179.8 | -14.81 |
| | PW91PW91 | 2.458 | 175.1 | -12.94 |
| 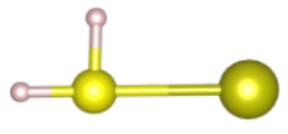 | PBE0 | 3.947 | 32.6 | -0.24 |
| | CAM-B3LYP | 4.272 | 21.2 | -0.07 |
| | PBE | 3.840 | 35.9 | -0.33 |
| | B3LYP | 7.167 | 5.7 | 0.13 |
| | PW91P86 | 3.667 | 44.0 | -0.94 |
| | PW91PW91 | 3.790 | 37.0 | -0.51 |
| 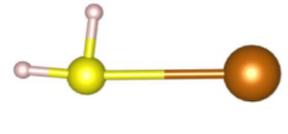 | PBE0 | 3.93 | 21.8 | -0.081 |
| | CAM-B3LYP | 4.317 | 16.8 | 0.089 |
| | PBE | | | |
| | B3LYP | | | |
| | PW91P86 | 3.525 | 43.7 | -0.95 |
| | PW91PW91 | 3.711 | 28.1 | -0.44 |
| 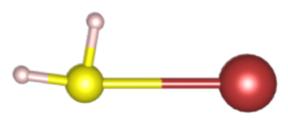 | PBE0 | 4.022 | 31.7 | -0.18 |
| | CAM-B3LYP | 4.214 | 22.7 | 0.12 |
| | PBE | 3.752 | 28.2 | -0.16 |
| | B3LYP | 4.677 | 19.2 | 0.19 |
| | PW91P86 | 3.496 | 45.3 | -0.91 |
| | PW91PW91 | 3.711 | 27.9 | -0.36 |

When there are more metallic atoms in the clusters, the adhesion between H$_2$S molecules and metallic cores occurs at gold sites or non-gold sites. As shown in **Error! Reference source not found.**, we found similar trends among bond lengths, vibrational frequencies, and binding energies: as the dopant increases from Zn to Hg, the vibrational frequency of S-Au and the binding energies decrease while the bond length increase. Thus, the dopant reduced the binding between H$_2$S and gold but the reduction decreases as the atomic radius increases. But with more gold atoms in the cores, such reduction becomes less significant.

## NBO analysis with natural charges of different molecules

*Table S3 Natural electron configuration and natural charge for H$_2$S-ZnAu$_n$, H$_2$S-CdAu$_n$, and H$_2$S-HgAu$_n$.(n=0-2) clusters.*

| | | H$_2$S-ZnAu$_n$ | | | H$_2$S-CdAu$_n$ | | | H$_2$S-HgAu$_n$ | |
|---|---|---|---|---|---|---|---|---|---|
| n | atom | natural electron configuration | natural charge | atom | natural electron configuration | natural charge | atom | natural electron configuration | natural charge |
| 0 | S | $3s^{1.76}3p^{4.55}3d^{0.03}4p^{0.02}$ | -0.352 | S | $3s^{1.76}3p^{4.57}3d^{0.03}4p^{0.02}$ | -0.374 | S | $3s^{1.76}3p^{4.56}3d^{0.03}4p^{0.02}$ | -0.369 |
| | H | $1s^{0.82}$ | 0.181 | H | $1s^{0.82}$ | 0.176 | H | $1s^{0.82}$ | 0.178 |
| | H | $1s^{0.82}$ | 0.181 | H | $1s^{0.82}$ | 0.181 | H | $1s^{0.82}$ | 0.181 |
| | Zn | $4s^{2.00}3d^{10.00}4p^{0.01}$ | -0.009 | Cd | $5s^{1.98}4d^{10.00}$ | 0.016 | Hg | $6s^{1.98}5d^{10.00}$ | 0.011 |
| 1 | S | $3s^{1.75}3p^{4.50}3d^{0.04}4p^{0.02}$ | -0.311 | S | $3s^{1.76}3p^{4.52}3d^{0.03}4p^{0.02}$ | -0.330 | S | $3s^{1.76}3p^{4.50}3d^{0.03}4p^{0.02}$ | -0.306 |
| | H | $1s^{0.82}$ | 0.182 | H | $1s^{0.84}$ | 0.156 | H | $1s^{0.81}$ | 0.189 |
| | H | $1s^{0.81}$ | 0.189 | H | $1s^{0.82}$ | 0.183 | H | $1s^{0.81}$ | 0.189 |

| | Zn | $4s^{1.48}3d^{9.99}4p^{0.24}$ | 0.278 | Cd | $5s^{1.49}4d^{9.99}5p^{0.18}$ | 0.326 | Hg | $6s^{1.73}5d^{9.98}6p^{0.11}$ | 0.162 |
|---|---|---|---|---|---|---|---|---|---|
| 2 | S | $3s^{1.73}3p^{4.46}3d^{0.04}4p^{0.01}5p^{0.01}$ | -0.248 | S | $3s^{1.74}3p^{4.49}3d^{0.03}4p^{0.01}5p^{0.01}$ | -0.279 | S | $3s^{1.75}3p^{4.53}3d^{0.02}4p^{0.01}5p^{0.01}$ | -0.338 |
| | H | $1s^{0.79}$ | 0.205 | H | $1s^{0.80}$ | 0.200 | H | $1s^{0.82}$ | 0.178 |
| | H | $1s^{0.79}$ | 0.205 | H | $1s^{0.80}$ | 0.200 | H | $1s^{0.86}$ | 0.142 |
| | Zn | $4s^{1.11}3d^{9.99}4p^{0.19}5p^{0.43}$ | 0.271 | Cd | $5s^{1.14}4d^{9.99}5p^{0.49}6p^{0.01}$ | 0.367 | Hg | $6s^{1.70}5d^{9.98}6p^{0.02}7s^{0.01}7p^{0.01}$ | 0.269 |

## NBO analysis of different molecules

*Table S4 NBO analysis results (H2S-MAun)*

| | H$_2$S-ZnAu$_n$ | | | H$_2$S-CdAu$_n$ | | | H$_2$S-HgAu$_n$ | | | H$_2$S-AgAu$_n$ | | |
|---|---|---|---|---|---|---|---|---|---|---|---|---|
| n | atom | natural charge | partition | atom | natural charge | partition | atom | natural charge | partition | atom | natural charge | partition |
| 2 | S | -0.248 | ligand | S | -0.279 | ligand | S | -0.338 | ligand | S | -0.243 | ligand |
| | H | 0.205 | | H | 0.200 | | H | 0.178 | | H | 0.207 | |
| | H | 0.205 | | H | 0.200 | | H | 0.142 | | H | 0.207 | |
| | Zn | 0.271 | | Cd | 0.367 | | Hg | 0.269 | | Ag | 0.178 | |
| | Au | -0.217 | core | Au | -0.244 | core | Au | -0.229 | core | Au | -0.175 | core |
| | Au | -0.217 | | Au | -0.244 | | Au | -0.023 | | Au | -0.175 | |
| 3 | S | -0.217 | ligand | S | -0.247 | ligand | S | | | S | -0.221 | ligand |
| | H | 0.194 | | H | 0.187 | | H | | | H | 0.194 | |
| | H | 0.194 | | H | 0.184 | | H | | | H | 0.194 | |
| | Zn | 0.100 | | Cd | 0.235 | | Hg | | | Ag | 0.177 | |
| | Au | -0.086 | core | Au | -0.152 | core | Au | | | Au | -0.012 | core |
| | Au | -0.086 | | Au | -0.112 | | Au | | | Au | -0.165 | |
| | Au | -0.100 | | Au | -0.095 | | Au | | | Au | -0.165 | |
| 4 | S | -0.190 | ligand | S | -0.237 | ligand | S | -0.277 | ligand | S | -0.216 | ligand |
| | H | 0.195 | | H | 0.180 | | H | 0.187 | | H | 0.190 | |
| | H | 0.195 | | H | 0.188 | | H | 0.174 | | H | 0.190 | |
| | Zn | -0.139 | | Cd | 0.346 | | Hg | 0.191 | | Ag | 0.050 | |
| | Au | -0.033 | core | Au | -0.052 | core | Au | -0.202 | core | Au | -0.012 | core |
| | Au | -0.033 | | Au | -0.139 | | Au | 0.035 | ligand | Au | -0.012 | |
| | Au | 0.003 | ligand | Au | -0.075 | | Au | -0.093 | core | Au | -0.095 | |
| | Au | 0.003 | | Au | -0.211 | | Au | -0.015 | | Au | -0.095 | |
| 6 | S | | | S | -0.212 | ligand | S | | | S | -0.228 | ligand |
| | H | | | H | 0.197 | | H | | | H | 0.188 | |
| | H | | | H | 0.197 | | H | | | H | 0.189 | |
| | Zn | | | Cd | 0.063 | | Hg | | | Ag | 0.100 | |
| | Au | | | Au | -0.043 | core | Au | | | Au | -0.040 | core |
| | Au | | | Au | 0.023 | ligand | Au | | | Au | 0.063 | ligand |
| | Au | | | Au | -0.026 | core | Au | | | Au | -0.049 | core |
| | Au | | | Au | 0.094 | ligand | Au | | | Au | -0.304 | |
| | Au | | | Au | -0.250 | | Au | | | Au | 0.026 | |
| | Au | | | Au | -0.043 | core | Au | | | Au | 0.055 | ligand |

| | H$_2$S-CuAu$_n$ | | | H$_2$S-NiAu$_n$ | | | H$_2$S-PdAu$_n$ | | | H$_2$S-PtAu$_n$ | | |
|---|---|---|---|---|---|---|---|---|---|---|---|---|
| n | atom | natural charge | core/ ligand | atom | natural charge | core/ ligand | atom | natural charge | core/ ligand | atom | natural charge | core/ ligand |
| | S | -0.226 | | S | -0.201 | | S | -0.200 | | S | -0.117 | |

| n | atom | natural charge | core/ligand | atom | natural charge | core/ligand | atom | natural charge | core/ligand | atom | natural charge | core/ligand |
|---|---|---|---|---|---|---|---|---|---|---|---|---|
| 2 | H | 0.201 | ligand | H | 0.167 | ligand | H | 0.178 | ligand | H | 0.176 | ligand |
|   | H | 0.201 |  | H | 0.167 |  | H | 0.178 |  | H | 0.176 |  |
|   | Cu | 0.075 |  | Ni | -0.169 |  | Pd | -0.224 | core | Pt | -0.419 | core |
|   | Au | -0.125 |  | Au | -0.055 | core | Au | 0.098 | ligand | Au | 0.189 | ligand |
|   | Au | -0.125 | core | Au | 0.091 | ligand | Au | -0.030 | core | Au | -0.004 | core |
| 3 | S | -0.191 |  | S |  |  | S |  |  | S |  |  |
|   | H | 0.192 | ligand | H |  |  | H |  |  | H |  |  |
|   | H | 0.192 |  | H |  |  | H |  |  | H |  |  |
|   | Cu | -0.031 |  | Ni |  |  | Pd |  |  | Pt |  |  |
|   | Au | -0.115 | core | Au |  |  | Au |  |  | Au |  |  |
|   | Au | 0.069 | ligand | Au |  |  | Au |  |  | Au |  |  |
|   | Au | -0.115 | core | Au |  |  | Au |  |  | Au |  |  |
| 4 | S | -0.180 |  | S |  |  | S | -0.077 |  | S |  |  |
|   | H | 0.190 | ligand | H |  |  | H | 0.177 | ligand | H |  |  |
|   | H | 0.193 |  | H |  |  | H | 0.177 |  | H |  |  |
|   | Cu | -0.203 |  | Ni |  |  | Pd | -0.542 | core | Pt |  |  |
|   | Au | -0.045 | core | Au |  |  | Au | 0.073 |  | Au |  |  |
|   | Au | 0.045 |  | Au |  |  | Au | 0.059 | core | Au |  |  |
|   | Au | 0.047 | ligand | Au |  |  | Au | 0.059 |  | Au |  |  |
|   | Au | -0.047 | core | Au |  |  | Au | 0.073 |  | Au |  |  |
| 6 | S | -0.166 |  | S |  |  | S | -0.133 |  | S | -0.036 |  |
|   | H | 0.193 | ligand | H |  |  | H | 0.176 | ligand | H | 0.166 | ligand |
|   | H | 0.192 |  | H |  |  | H | 0.176 |  | H | 0.166 |  |
|   | Cu | -0.552 | core | Ni |  |  | Pd | -0.598 | core | Pt | -0.817 |  |
|   | Au | 0.097 |  | Au |  |  | Au | 0.134 | ligand | Au | -0.033 | core |
|   | Au | 0.050 | ligand | Au |  |  | Au | -0.055 | core | Au | 0.145 |  |
|   | Au | 0.039 |  | Au |  |  | Au | 0.063 |  | Au | 0.187 | ligand |
|   | Au | 0.074 |  | Au |  |  | Au | 0.147 | ligand | Au | 0.068 |  |
|   | Au | -0.013 | core | Au |  |  | Au | 0.147 |  | Au | 0.187 |  |
|   | Au | 0.085 | ligand | Au |  |  | Au | -0.055 | core | Au | -0.033 | core |

*Table S5 NBO analysis results (H2S-AunM)*

| | H$_2$S-Au$_n$Zn | | | H$_2$S-Au$_n$Cd | | | H$_2$S-Au$_n$Hg | | | H$_2$S-Au$_n$Ni | | |
|---|---|---|---|---|---|---|---|---|---|---|---|---|
| n | atom | natural charge | core/ligand | atom | natural charge | core/ligand | atom | natural charge | core/ligand | atom | natural charge | core/ligand |
| 2 | S | -0.155 |  | S | -0.159 |  | S | -0.162 |  | S | -0.191 |  |
|   | H | 0.202 | ligand | H | 0.201 | ligand | H | 0.203 | ligand | H | 0.187 | ligand |
|   | H | 0.204 |  | H | 0.204 |  | H | 0.204 |  | H | 0.186 |  |
|   | Au | -0.238 |  | Au | -0.205 |  | Au | -0.145 |  | Au | -0.140 |  |
|   | Au | -0.177 | core | Au | -0.189 | core | Au | -0.177 | core | Au | -0.166 | core |
|   | Zn | 0.164 | ligand | Cd | 0.148 | ligand | Hg | 0.076 | ligand | Ni | 0.124 | ligand |

| | H$_2$S-Au$_n$Pd | | | H$_2$S-Au$_n$Pt | | | H$_2$S-Au$_n$Cu | | | H$_2$S-Au$_n$Ag | | |
|---|---|---|---|---|---|---|---|---|---|---|---|---|
| n | atom | natural charge | core/ligand | atom | natural charge | core/ligand | atom | natural charge | core/ligand | atom | natural charge | core/ligand |
| | S | -0.183 |  | S | -0.167 |  | S | -0.179 |  | S | -0.173 |  |
| | H | 0.198 | ligand | H | 0.198 | ligand | H | 0.202 | ligand | H | 0.202 | ligand |

| | | | | | | | | | | | |
|---|---|---|---|---|---|---|---|---|---|---|---|
| 2 | H | 0.196 | | H | 0.196 | | H | 0.197 | | H | 0.198 | |
| | Au | -0.097 | | Au | -0.029 | core | Au | -0.158 | | Au | -0.146 | |
| | Au | -0.097 | core | Au | 0.071 | ligand | Au | -0.225 | core | Au | -0.203 | core |
| | Pd | -0.017 | | Pt | -0.268 | core | Cu | 0.163 | ligand | Ag | 0.122 | ligand |
| 4 | | | | | | | S | -0.158 | | S | -0.162 | |
| | | | | | | | H | 0.192 | ligand | H | 0.188 | ligand |
| | | | | | | | H | 0.188 | | H | 0.190 | |
| | | | | | | | Au | -0.266 | | Au | -0.312 | |
| | | | | | | | Au | -0.091 | core | Au | -0.049 | core |
| | | | | | | | Au | -0.064 | | Au | -0.127 | |
| | | | | | | | Au | -0.026 | | Au | -0.057 | |
| | | | | | | | Cu | 0.225 | ligand | Ag | 0.328 | ligand |
| 6 | | | | | | | S | -0.195 | | S | -0.222 | |
| | | | | | | | H | 0.193 | ligand | H | 0.194 | ligand |
| | | | | | | | H | 0.193 | | H | 0.194 | |
| | | | | | | | Au | -0.497 | core | Au | 0.056 | |
| | | | | | | | Au | 0.068 | | Au | 0.056 | |
| | | | | | | | Au | 0.068 | | Au | -0.002 | |
| | | | | | | | Au | 0.030 | core | Au | -0.002 | core |
| | | | | | | | Au | 0.030 | | Au | 0.068 | |
| | | | | | | | Au | 0.068 | | Au | -0.548 | |
| | | | | | | | Cu | 0.043 | | Ag | 0.208 | ligand |